\documentclass[twocolumn]{jpsj3}
\usepackage{txfonts}

\usepackage[dvipdfmx]{color}
\usepackage[dvipdfmx]{graphicx}
\usepackage{dcolumn}
\usepackage{bm}
\usepackage[normalem]{ulem} 

\title{
  Reply to Comments on ``Universal and Non-Universal Correction Terms 
  of Bose Gases in Dilute Region: A Quantum Monte Carlo Study''
}

\author{
  Akiko Masaki-Kato$^{1,2}$,
  Yuichi Motoyama$^3$ and
  Naoki Kawashima$^3$\thanks{kawashima@issp.u-tokyo.ac.jp}
}
\inst{
  $^1$Computational Condensed Matter Physics Laboratory, RIKEN, Wako,
  Saitama 351-0198, Japan \\
  $^2$Computational Materials Science Research Team,
  RIKEN Center for Computational Science (R-CCS),  Kobe, Hyogo 650-0047, Japan \\
  $^3$The Institute for Solid State Physics, The University of Tokyo, Chiba, Japan 277-8581
} 

\begin{document}
\maketitle

In the recent paper \cite{MasakiKato} (referred to as ``our previous paper'' hereafter), we presented some results of our numerical calculation of the hard-core bosonic Hubbard model, and by comparing them with the analytic predictions, such as the energy formula,
\begin{eqnarray}
  \frac{E}{N}
  &=& \frac{2\pi\hbar^2}{ma_s^2}\times \tilde n \times \left[ 1+
    \left(\frac{128}{15\sqrt{\pi}} \right)\sqrt{\tilde n}\right.\nonumber\\
  &+&
    \left.\frac{8(4\pi-3\sqrt{3})}{3}\tilde n\log(\tilde n)+ c_3 \tilde n 
    + o(\tilde n) \right],
\label{LHY}
\end{eqnarray}
where $\tilde n \equiv n a_s^3$,
we estimated the $s$-wave scattering length $a_s$ of this model in addition to the amplitude of the leading non-universal correction term, which we referred to as $c_3$. As we noted at the end of our previous paper, F\'elix Werner \cite{Werner}, quoting Castin\cite{Castin}, brought our attention to the fact that the standard scattering theory reduces the $s$-wave scattering length to a three-dimensional wave-number integral. The numerical estimate of the integral is 0.314870..., in good agreement with our Monte Carlo estimate 0.316\,(2). Furthermore, Adam Ran\c{c}on\cite{Rancon} pointed out that this wave-number integral actually has a simple expression in terms of the Gamma function. He further showed that a careful spin-wave analysis yields the formula (\ref{LHY}) up to the second term in addition to a non-universal correction term. The non-universal correction term derived from the spin wave theory may not be equal to but yields a contribution of a few percent of the  $c_3$ term estimated in our previous paper. He suggested that we may improve the estimate of the $c_3$ term by using the exact value of the $s$-wave scattering length, and further quantitative comparison between the Monte Carlo results and the spin-wave approximation. 

\begin{table}[b]
    \centering
    \begin{tabular}{ccc|cl}
    \hline
          & \hspace{1mm} $L$ \hspace{1mm} & \hspace{4mm} $\beta$ \hspace{4mm} & & \hspace{3mm} $c_3$ \hspace{3mm} \\
    \hline
         & 32 &  64 & & 119.3\,(7) \\
         & 32 &  96 & & 123\,(11) \\
         & 48 &  96 & & 119.9\,(10) \\
    \hline
    \end{tabular}
    \caption{The estimate of the coefficient $c_3$ for three sets of the system sizes with the exact value of the $s$-wave scattering length provided by Ran\c{c}on\cite{Rancon}.}
    \label{tab:my_label}
\end{table}
Both suggestions are quite reasonable and interesting, although we cannot answer the second suggestion promptly and have to leave it for a future work. Here, we present the improved estimate of the $c_3$-term. 
As in our previous paper, we use the three data sets of the energy estimates of the hard-core Bose-Hubbard model, as a function of the chemical potential. Each data set is characterized by the inverse temperature $\beta$ and the linear size $L$.
The size dependence does not seem significant beyond the statistical error, which is in agreement with the validity of the exact value for $a_s$. We here quote the result of $(L,\beta)=(32,64)$ as our estimate of $c_3$:
\begin{equation}
    c_3 = 119.3\,(7).
\end{equation}
As F\'elix Werner\cite{Werner} pointed out, 
this value is, interestingly, not far from the estimates for other systems,
such as $c_3 =  167.319 69\,(6)$ for the hard-spheres in continuous space\cite{Tan},
${\rm Re}(c_3) \sim 141$ for the dilute Bose gas with resonant interactions \cite{Braaten}
and the experimental estimate $c_3 = 140\,(20)$ for the resonant Fermi gas in the BEC limit\cite{Navon},
in spite that $c_3$ is non-universal and therefore may in principle depend on the details of the scattering potential and whether the space is continuous or discrete. 



\begin{thebibliography}{99}
\bibitem{MasakiKato}
  A. Masaki-Kato, Y. Motoyama, and N. Kawashima, J. Phys. Soc. Jpn. 91, 024001 (2022).
\bibitem{Werner}
  F. Werner, private communication.
\bibitem{Castin}
  Y. Castin, "Basic tools for degenerate Fermi gases," 
  in Proceedings of the International School of Physics “Enrico Fermi”, 164, 289 (2007).
\bibitem{Rancon}
  Adam Ran\c{c}on, to appear in J. Phys. Soc. Jpn. 
\bibitem{Tan}
  S. Tan, Phys. Rev. A 78, 013636 (2008).
\bibitem{Braaten}
  E. Braaten, H. W. Hammer, and T. Mehen, Phys. Rev. Lett. 88, 040401 (2002).
\bibitem{Navon}
  N. Navon, S. Nascimbne, F. Chevy, and C. Salomon, Science 328, 729 (2010).
\end{thebibliography}
\end{document}